\documentclass[9pt,conference,a4paper]{IEEEtran}
\ifCLASSINFOpdf
\else
\fi
\usepackage{amssymb}
\usepackage{bm}
\usepackage[french,english]{babel}
\usepackage{color}
\usepackage{amsmath}
\usepackage[]{graphicx}

\begin{document}

\title{Sparsity averaging for radio-interferometric imaging}

\author{%
\IEEEauthorblockN{
Rafael E. Carrillo\IEEEauthorrefmark{1}, 
Jason D. McEwen\IEEEauthorrefmark{2}, 
and Yves Wiaux\IEEEauthorrefmark{1}\IEEEauthorrefmark{3}\IEEEauthorrefmark{4}}
\IEEEauthorblockA{\IEEEauthorrefmark{1} 
Institute of Electrical Engineering, Ecole Polytechnique F{\'e}d{\'e}rale de Lausanne (EPFL),
      CH-1015 Lausanne, Switzerland.}
\IEEEauthorblockA{\IEEEauthorrefmark{2}
Department of Physics and Astronomy, University College London, London WC1E 6BT, UK.}
\IEEEauthorblockA{\IEEEauthorrefmark{3}
Department of Radiology and Medical Informatics, University of Geneva (UniGE), 
      CH-1211 Geneva, Switzerland.}
\IEEEauthorblockA{\IEEEauthorrefmark{4}
Department of Radiology, Lausanne University Hospital (CHUV), CH-1011 Lausanne, Switzerland.
}
}

\maketitle

We propose a novel regularization method for compressive imaging in the context of the compressed sensing (CS) theory with coherent and redundant dictionaries \cite{candes10}. Natural images are often complicated and several types of structures can be present at once. It is well known that piecewise smooth images exhibit gradient sparsity, and that images with extended structures are better encapsulated in wavelet frames. Therefore, we here conjecture that promoting average sparsity or compressibility over multiple frames rather than single frames is an extremely powerful regularization prior. Define $\bm{x}\in\mathbb{R}^{N}$ to be the image of interest. We propose using a dictionary composed of a concatenation of $q$ frames, i.e.
\begin{equation}
\mathsf{\Psi}=\frac{1}{\sqrt{q}}[\mathsf{\Psi}_1, \mathsf{\Psi}_2, \ldots, \mathsf{\Psi}_q],
\end{equation}
and an analysis $\ell_0$ prior, $\|\mathsf{\Psi}^{\dagger}\bm{x}\|_{0} $, to promote this average sparsity.

Note on a theoretical level that a single signal cannot be arbitrarily sparse simultaneously in any pair of frames. For example, a signal extremely sparse in the Dirac basis is completely spread in the Fourier basis. As discussed in \cite{carrillo12b}, each frame, $\mathsf{\Psi}_i$, should be highly coherent with the other frames in order to have a sparse representation for the signal. The concatenation of the Dirac basis and the first eight orthonormal Daubechies wavelet bases (Db1-Db8) represents a natural choice in the imaging context. The first Daubechies wavelet basis, Db1, is the Haar wavelet basis and, in particular, can be used as an alternative to gradient sparsity (usually imposed by a total variation (TV) prior) to promote piecewise smooth signals. The Db2-Db8 bases are coherent with Haar and Dirac while providing smoother decompositions.

The proposed approach is defined on the basis of the following problem:
\begin{equation}\label{delta}
\min_{\bar{\bm{x}}\in\mathbb{R}^{N}_{+}}\|\mathsf{\Psi}^{\dagger}\bar{\bm{x}}\|_{0}\textnormal{  subject to  }\| \bm{y}-\mathsf{\Phi}\bar{\bm{x}}\|_{2}\leq\epsilon,
\end{equation}
where the matrix $\mathsf{\Phi}\in\mathbb{C}^{M\times N}$ identifies the measurement operator, $\bm{y}\in\mathbb{C}^{M}$ identifies the measurement vector and $\epsilon$ is an upper bound on the $\ell_2$ norm of the residual noise. The constraint $\bar{\bm{x}}\in\mathbb{R}^{N}_{+}$ represents the positivity prior on $\bm{x}$. Practically, the algorithm uses a reweighted approach to approximate $\ell_0$ minimization by solving a sequence of weighted $\ell_1$ problems. The associated reconstruction algorithm is dubbed Sparsity Averaging Reweighted Analysis (SARA). See \cite{carrillo12b} for a detailed description of the algorithm. This problem offers great versatility since one only needs to model correctly the measurement operator to allow reconstruction from different sensing modalities. Results comparing SARA to state-of-the-art reconstruction methods for random Gaussian, spread spectrum, and random discrete Fourier sampling are reported in \cite{carrillo12b, carrillo12a}. The results demonstrate that SARA outperforms benchmark methods in all cases.

\begin{figure}[h]
    \centering
    \includegraphics[trim = 3.4cm 1.1cm 2.2cm 1.0cm, clip, keepaspectratio, width = 2.8cm]{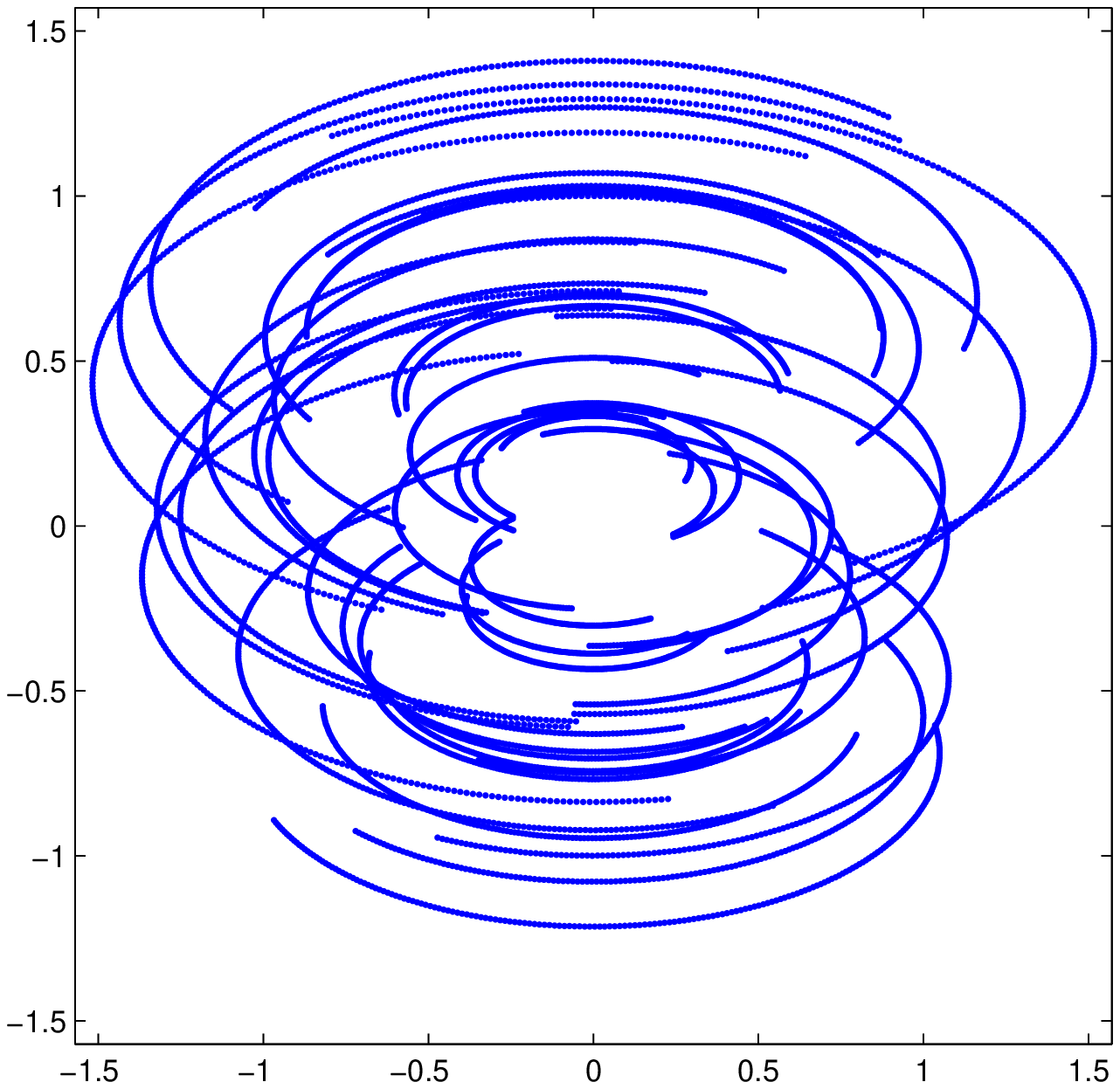}
    \includegraphics[trim = 3.4cm 1.1cm 2.2cm 1.0cm, clip, keepaspectratio, width = 2.8cm]{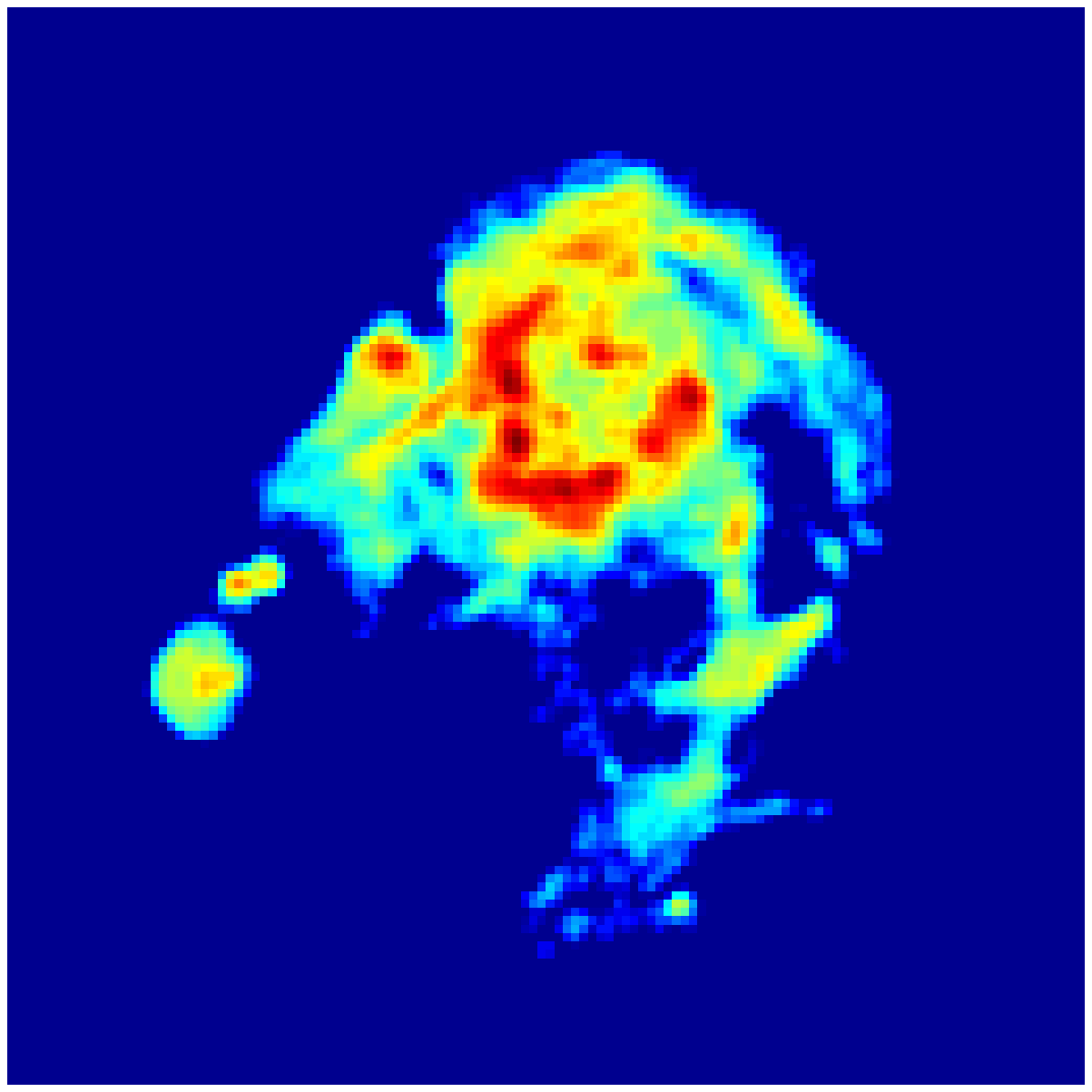}
    \includegraphics[trim = 3.4cm 1.1cm 2.2cm 1.0cm, clip, keepaspectratio, width = 2.8cm]{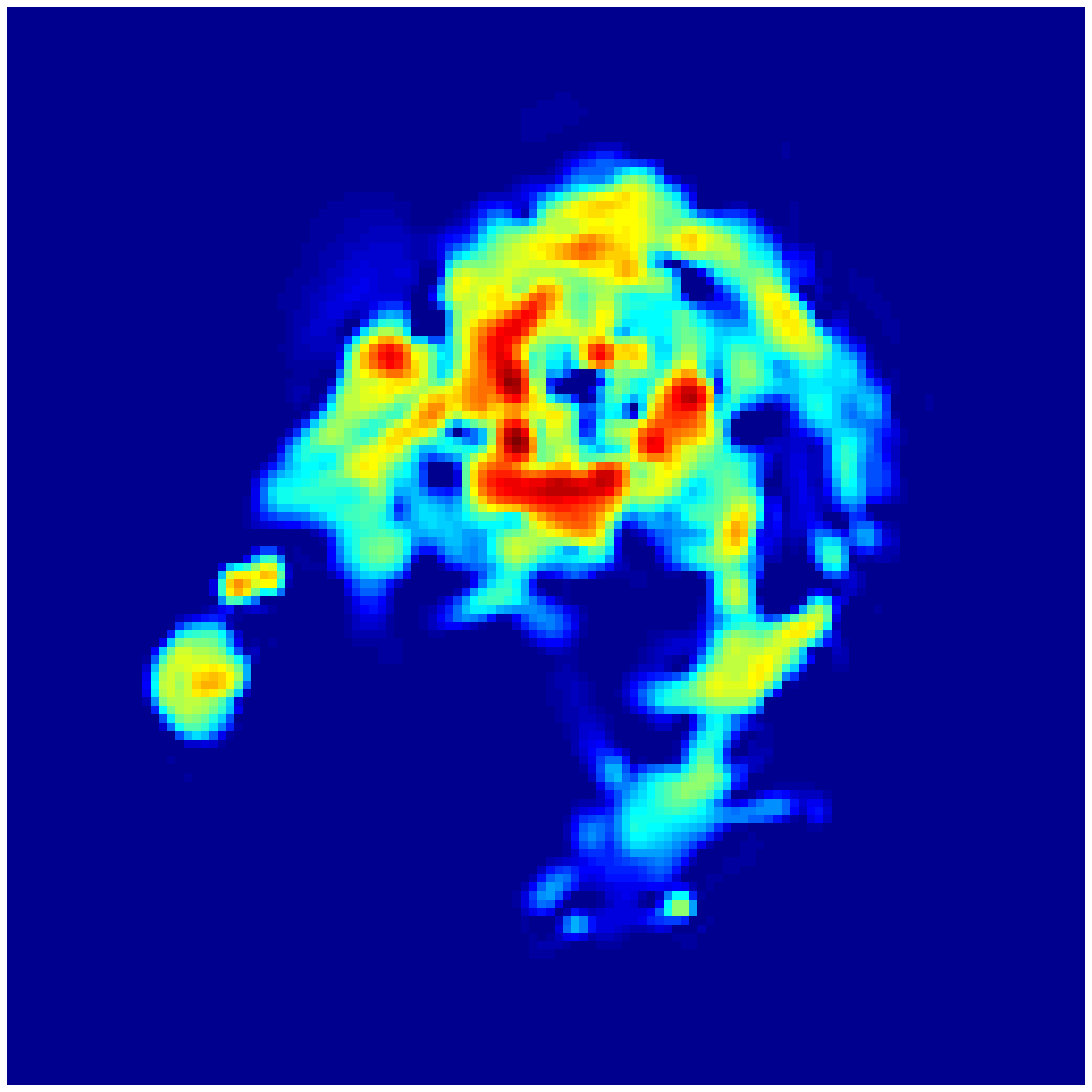}
    \includegraphics[trim = 3.4cm 1.1cm 2.2cm 1.0cm, clip, keepaspectratio, width = 2.8cm]{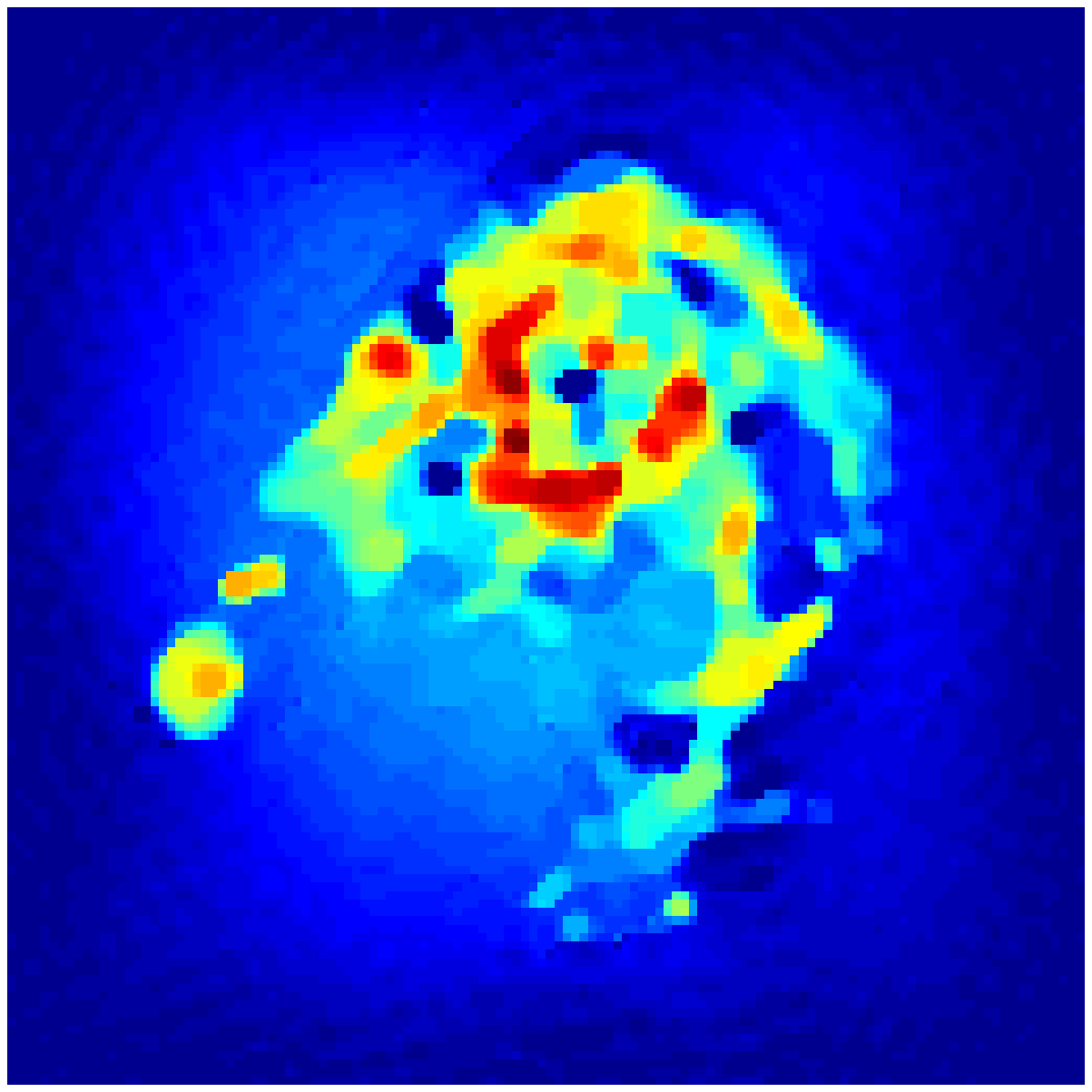}
    \includegraphics[trim = 3.4cm 1.1cm 2.2cm 1.0cm, clip, keepaspectratio, width = 2.8cm]{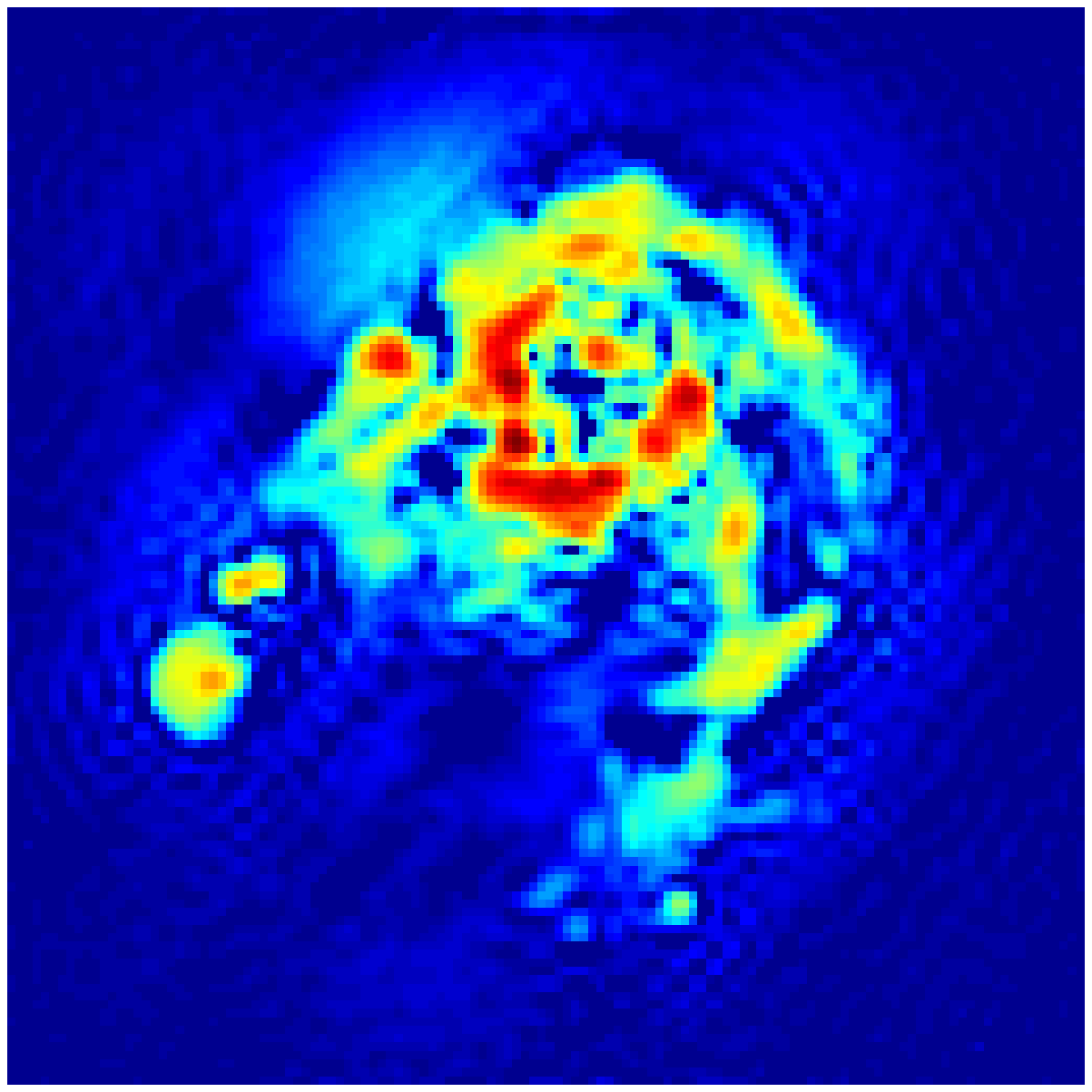}
    \includegraphics[trim = 3.4cm 1.1cm 2.2cm 1.0cm, clip, keepaspectratio, width = 2.8cm]{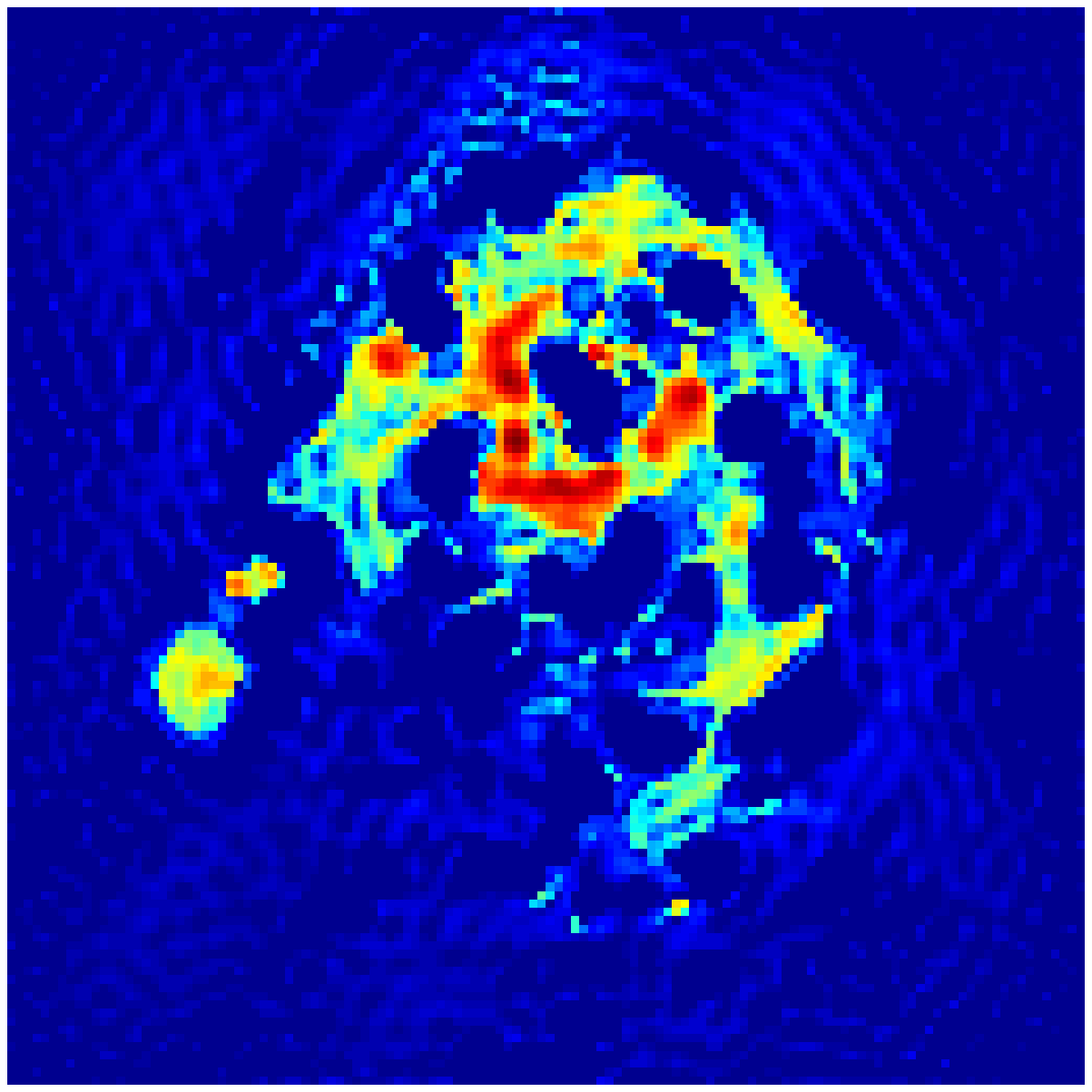}
\caption{From left to right, top to bottom: Fourier sampling profile, original test image and reconstructions for SARA (13.35~dB), TV (11.44~dB), BPDb8 (11.13~dB) and BP (8.19~dB) in $\log_{10}$ scale.}
\label{fig:2}
\end{figure}

In the case of radio-interferometric (RI) imaging, under common assumptions, the measurement equation for aperture synthesis provides incomplete Fourier sampling of the image of interest. Since the measured visibilities provide continuous samples of the frequency plane, an interpolation operator needs to be included in $\mathsf{\Phi}$ to model the map from a discrete frequency grid onto the continuous plane so that the FFT can be used. Direction dependent effects can also be included in the model as additional convolution kernels in the frequency plane. We here illustrate the performance of SARA in this field by recovering the well known test image of the M31 galaxy from simulated continuous visibilities affected by 30 dB of input noise, and using the Dirac-Db1-Db8 concatenation highlighted for $\mathsf{\Psi}$. For comparison, we also study a variety of sparsity-based image reconstruction algorithms, some of which were identified as providing similar performance as CLEAN and its multi-scale versions, which are state of the art in RI imaging. SARA is compared with the following minimization problems: (i) BP, constrained $\ell_1$-minimization in the Dirac basis (similar to CLEAN), (ii) BPDb8, constrained analysis-based $\ell_1$-minimization in the Db8 basis (similar to multi-scale CLEAN), and (iii) TV, constrained TV-minimization. Figure~\ref{fig:2} shows preliminary results of a reconstruction from a realistic radio telescope sampling pattern. SARA provides not only a drastic SNR increase but also a significant reduction of visual artifacts relative to all other methods. Note that all these algorithms are currently being implemented in a new package written in C named PURIFY.

\vspace{-0.4cm}
\bibliographystyle{IEEEtran}
\bibliography{bib}

\end{document}